\documentstyle[preprint,aps,epsf,floats]{revtex}    % PREPRINT STYLE
\def\lsim{\mathrel{\raise.2ex\hbox{$<$}\hskip-.8em\lower.9ex\hbox{$\sim$}}}
\def\gsim{\mathrel{\raise.2ex\hbox{$>$}\hskip-.8em\lower.9ex\hbox{$\sim$}}}
\begin{document}
\tighten
%\everymath={\displaystyle}
%
%%%%%%%%%%%%%%%%%%%%% TITLE PAGE %%%%%%%%%%%%%%%%%%%%%%%%%%%%%%%%%%%%%%%%%%%%%%
%
\preprint{
\font\fortssbx=cmssbx10 scaled \magstep2
\hbox to \hsize{
%\special{psfile=uwlogo.ps
%hscale=8000 vscale=8000
%hoffset=-12 voffset=-2}
%\hskip.5in \raise.1in\hbox{\fortssbx University of Wisconsin - Madison}
\hfill$\vtop{   \hbox{\bf TTP96-10}
                \hbox{\bf MADPH-96-935}
%                \hbox{\bf hep-ph/9511???\\}
                \hbox{March 1996}}$ }
}

\title{Jet Production in Deep Inelastic Scattering at HERA 
\footnotemark[1]              
}
\footnotetext[1]{Invited talk given at the Cracow Epiphany Conference
on Proton Structure, Krakow, January 5-6,1996;
presented by E.~Mirkes}
\author{Erwin Mirkes$^a$ and Dieter Zeppenfeld$^b$\\[3mm]}
\address{$^a$Institut f\"ur Theoretische Teilchenphysik, 
         Universit\"at Karlsruhe, D-76128 Karlsruhe, Germany\\[2mm]}
\address{
$^b$Department of Physics, University of Wisconsin, Madison, WI 53706, USA}
\maketitle
\begin{abstract}
Two-jet cross sections in deep inelastic scattering at HERA are calculated in 
next-to-leading order. 
The importance of higher order corrections and recombination scheme 
dependencies is studied for various jetalgorithms.
Some implications for the determination of 
$\alpha_s(\mu_R^2)$, the determination of the gluon density and
the associated forward jet production in the low $x$ regime at HERA
are briefly discussed.
\end{abstract}
%
%\pacs{PACS numbers: 13.35.+s, 11.30.Rd, 12.40.Vv}
%
\newpage
%
%%%%%%%%%%%%%%%%%% MAIN TEXT %%%%%%%%%%%%%%%%%%%%%%%%%%%%%%%%%%%%%%%%%%%%%%%
%
\narrowtext
\section{Introduction}
Deep inelastic scattering (DIS) at HERA is a copious source of 
multi-jet events. Typical two-jet cross sections\footnote{In the following 
the jet due to the beam remnant is not included in the number of jets.} 
are in the 100~pb to few nb range and thus provide sufficiently high 
statistics for precision QCD tests~\cite{exp_as}.  
Clearly, next-to-leading order (NLO) QCD corrections
are mandatory on the theoretical side for such tests.
Full NLO corrections for one and two-jet production cross sections
and distributions are now available and implemented in the 
$ep \rightarrow n$ jets event generator MEPJET,
which allows to analyze 
arbitrary jet definition schemes and 
general cuts in terms of parton 4-momenta\cite{mz1}.
A variety of topics can be studied with these tools. They include 
\begin{itemize}
\item The determination of 
      $\alpha_s(\mu_R^2)$ over a wide range of scales:
   The dijet cross section
   is proportional to $\alpha_s(\mu_R)$ at leading order (LO), thus suggesting 
   a direct measurement of the strong coupling constant. However, the
   LO calculation leaves the renormalization scale $\mu_R$ undetermined.
   The NLO corrections substantially reduce the renormalization and 
   factorization scale dependencies which are present in the LO calculations 
   and thus reliable cross section predictions in terms of $\alpha_s(m_Z)$
   (for a given set of parton distributions)
   are made possible.
\item The measurement of the 
      gluon density in the proton (via $\gamma g\to q\bar q$):
      The gluon density can only be indirectly constrained
      by an analysis of the structure  function $F_2$ at HERA
      \cite{exp_gluon}.
      The boson gluon fusion subprocess         
      dominates the two jet cross section
      at low $x$ and allows for a more direct measurement of the gluon density
      in this regime. A first LO  experimental analysis
      has been presented in \cite{h1_gluon}.
      NLO corrections reduce the factorization scale dependence
      in the LO calculation (due to the intital state collinear factorization,
      which introduces a mixture of the quark and gluon densities 
      according to the Altarelli-Parisi evolution)  
      and thus reliable cross section predictions in terms of the
      scale dependent  parton distributions are made possible.
 
\item The study of internal jet structure:
      NLO corrections in jet physics imply that 
      a jet (in a given jet definition scheme) may consist of two partons.
      Thus first sensitivity to the internal jet structure is obtained, 
      like dependence on the cone size or on recombination prescriptions.
      These studies are also important for reliable
      QCD studies (such as the $\alpha_s$ or gluon density determinations).
      The recombination dependence is only simulated at tree level 
      in the NLO calculation and  thus the dependence of the
      cross section on the recombination scheme is subject to
      potentially large higher oder corrections.

\item Associated 
      forward jet production in the low $x$ regime as a signal of BFKL
      dynamics:
      BFKL evolution \cite{bfkl} leads to a larger cross section for 
      events with a measured forward jet (in the proton
      direction) with transverse momentum $p_T^{lab}(j)$ 
      close to $Q$ than the DGLAP \cite{dglap} evolution.
      Clearly, next-to-leading order QCD corrections
      for fixed order QCD, with Altarelli-Parisi (DGLAP) evolution, 
      are mandatory on the theoretical side in order to establish a 
      signal for BFKL evolution in the data.

\item The determination of the polarized gluon structure function
       (via $\gamma g\to q\bar q$) in polarized electron on polarized proton
       scattering \cite{ziegler}: 
       The measurement of the polarized parton densities
       and in particular the polarized gluon density 
       from dijet production at a polarized
       electron proton machine at HERA energies would allow 
       to discriminate between the different pictures of the proton
       spin underlying these parametrizations.
       We will analyze these effects in a subsequent publication.
\end{itemize}
 Some theoretical aspects related to these studies are  discussed in the
following.

\section{Technical Matters and Jet Algorithms}
The goal of a versatile NLO calculation is to allow for an easy 
implementation of an arbitrary jet algorithm or to impose any kinematical
resolution and acceptance cuts on the final state particles. This is best 
achieved by performing all hard phase space integrals numerically, 
with a Monte Carlo integration technique. 
This approach also allows an investigation of the recombination scheme
dependence of the NLO jet cross sections. For dijet production at HERA such
a NLO Monte Carlo program is MEPJET.

The basic features of the calculation are described in \cite{mz1}
and we repeat only some of them here.
In Born approximation, the subprocesses  
$\gamma^\ast +q \rightarrow q + g$, 
$\gamma^\ast +\bar q \rightarrow \bar q + g$,  
and
$\gamma^\ast+g \rightarrow q + \bar{q}$  
contribute to the two-jet cross section. 
At ${\cal O}(\alpha_{s}^{2})$ the real emission corrections involve   
$\gamma^\ast+q \rightarrow q + g + g,\,
 \gamma^\ast+q \rightarrow q + \bar{q} + q,\,
 \gamma^\ast+g \rightarrow q + \bar{q} + g\,$
and analogous anti-quark initiated processes. The corresponding 
cross sections are calculated by 
numerically evaluating the tree level helicity amplitudes as given in 
Ref.~\cite{HZ}.
The tree level matrix elements are numerically checked against the
matrix elements in \cite{heraii,disjet}.
They need to be integrated over the entire phase space,
including the unresolved regions, where only two jets
are reconstructed according to a given jet definition scheme.  In order
to isolate the infrared as well as collinear divergencies associated with
these unresolved regions the resolution parameter $s_{min}$ is introduced.
This $s_{min}$ technique has already been successfully 
applied to  next-to-leading 
order calculations of jet cross sections in $e^+e^-$ annihilation and
in hadronic collisions~\cite{giele1,giele2}.
Soft and collinear approximations are used in the region where at least one 
pair of partons, including initial ones, has $s_{ij}=2p_i\cdot p_j<s_{min}$
and the soft and/or collinear final state parton is integrated over 
analytically. Factorizing the collinear initial state divergencies into the 
bare parton distribution functions and adding this soft+collinear part to the
${\cal{O}}(\alpha_s^2)$ virtual contributions 
for the 
$\gamma^\ast +q \rightarrow q + g$, 
$\gamma^\ast +\bar q \rightarrow \bar q + g$,  
and
$\gamma^\ast+g \rightarrow q + \bar{q}$  
subprocesses
gives a finite result for, effectively, 2-parton final states. In general 
this 2-parton contribution is negative and
grows logarithmically in magnitude as $s_{min}$ is decreased. This logarithmic 
growth is exactly cancelled by the increase in the 3 parton cross 
section, once $s_{min}$ is small enough for the approximations to be valid.

As mentioned before the 
collinear initial state divergencies are factorized into 
the bare parton densities introducing a dependence on the factorization 
scale $\mu_F$. In order to handle these singularities we follow 
Ref.~\cite{giele2} and use the technique of universal ``crossing
functions''.

The integration over the 3-parton phase space with $s_{ij}>s_{min}$
is done by Monte-Carlo techniques (without using
any approximations). 
In general, $s_{min}$ has to be chosen fairly small (below
$\leq 0.1$ GeV$^2$). Therefore, the effective 2-parton final state
(soft+collinear+virtual part) does not depend on the
recombination scheme. The essential benefit of the
Monte Carlo approach in MEPJET is that all hard phase space integrals 
over the region $s_{ij}>s_{min}$ are performed numerically. 
Since, at each phase space point, the parton 4-momenta 
are available, the program is flexible enough to implement 
arbitrary jet algorithms and
kinematical resolution and acceptance cuts.

For the numerical studies below, 
the standard set of parton distribution functions is MRS set 
D-' \cite{mrs}.
We employ the two loop formula for the strong coupling constant
%
%
%\begin{equation}
%\alpha_{s\,\overline{MS}}(\mu_R^2)= 
%\frac{12\pi}{(33-2n_{f})\ln(\mu_R^2/\Lambda^{2})}
%\left[1-\frac{6(153-19n_{f})}{(33-2n_{f})^{2}}
%\frac{\ln\ln(\mu_R^2/\Lambda^{2})}{\ln(\mu_R^2/\Lambda^{2})}\right]
%\label{asnlo}
%\end{equation}
%
%
with %${\Lambda_{\overline{MS}}}={\Lambda_{\overline{MS}}^{(4)}}=230$ MeV, 
${\Lambda_{\overline{MS}}^{(4)}}=230$ MeV, 
which is the value from the parton distribution functions.
The value of $\alpha_s$ is matched at the thresholds $\mu_R=m_q$ and the
number of flavors is fixed to $n_f=5$ throughout, {\it i.e.} gluons are 
allowed to split into five flavors of massless quarks.
Unless stated otherwise, the renormalization scale  and
the factorization scale  are set to
$\mu_R=\mu_F= 1/2\,\sum_i \,p_T^B(i)$, where $p_T^B(i)$ denotes the magnitude
of the transverse momentum of parton $i$ in the Breit frame.
A running QED fine structure constant $\alpha(Q^2)$ is used.
The lepton and hadron beam energies are 27.5 and 820 GeV, respectively.
A  minimal set of kinematical cuts is imposed on the initial virtual photon 
and on the final state electron and jets. We require 
40~GeV$^2<Q^2<2500$ GeV$^2$,
$0.04 < y < 1$, an energy cut of $E(e^\prime)>10$~GeV on the scattered 
electron, and a cut on the pseudo-rapidity $\eta=-\ln\tan(\theta/2)$
of the scattered lepton and jets of $|\eta|<3.5$. In addition jets 
must have transverse momenta of at least 2~GeV in both the lab and the 
Breit frame.

Within these general cuts four different jet definition schemes are 
considered for which we have chosen parameters such as to give similar
LO cross sections. 
Note however, that the phase space region for the accepted  dijet events
depends on the jetalgorithm and thus somewhat different event sets would be 
considered in an actual experiment.
Unless stated otherwise, and for all
jet algorithms, we use the $E$-scheme to recombine partons, i.e.
the cluster momentum is taken as $p_i+p_j$, the sum of the 4-momenta
of partons $i$ and $j$, if these
are unresolved according to a given jet definition scheme.
\begin{itemize}
\item[1)]  $W$-scheme:\\
In the $W$-scheme the invariant mass squared, 
$s_{ij}=(p_i+p_j)^2$, is calculated for each pair of final state 
particles (including the proton remnant) \cite{heraii}.
If the pair with the smallest invariant mass squared is below $y_{cut}W^2$,
the pair is clustered according to a recombination scheme. 
%Unless stated otherwise, and for all
%jet algorithms, we use the $E$-scheme to recombine partons, i.e.
%the cluster momentum is taken as $p_i+p_j$. 
This process 
is repeated until all invariant masses are above $y_{cut} W^2$.        
The  resolution parameter $y_{cut}$ is fixed to $0.02$.

\item[2)] JADE-scheme:\\ 
The experimental analyses in \cite{exp_as} are  based 
on a variant of the $W$-scheme, the ``JADE'' algorithm \cite{jade}. It is
obtained from the $W$-scheme by replacing the invariant
definition $s_{ij}=(p_i+p_j)^2$ by 
$M_{ij}^2=2E_iE_j(1-\cos\theta_{ij})$, where all quantities are defined 
in the laboratory frame. Neglecting the explicit mass terms $p_i^2$ 
and $p_j^2$ in the definition of $M_{ij}^2$ causes substantial
differences in jet cross sections between the $W$ and the JADE scheme.

\item[3)] cone scheme:\\
In the cone algorithm (which is defined in the laboratory frame) the distance 
$\Delta R=\sqrt{(\Delta\eta)^2+(\Delta\phi)^2}$ between two partons 
decides whether they should be recombined to a single jet. Here the variables 
are the pseudo-rapidity $\eta$ and the azimuthal angle $\phi$. We 
recombine partons with $\Delta R<1$.
Furthermore, a cut on the jet transverse momenta of $p_T(j)>5$~GeV in the lab 
frame is imposed in addition to the 2 GeV Breit frame cut.

\item[4)] $k_T$ scheme:\\
For the $k_T$ algorithm (which is implemented in the Breit frame), 
we follow the description introduced
in Ref.~\cite{kt}. The hard scattering scale $E_T^2$ is
fixed to 40 GeV$^2$ and $y_{cut}=1$ is the resolution parameter 
for resolving the macro-jets. In addition, jets are required to have a minimal 
transverse momentum of 5 GeV in the Breit frame.
\end{itemize}

\setlength{\unitlength}{0.7mm}
\begin{figure}[hbt]               \vspace*{-5cm}
\begin{picture}(150,165)(-30,1)
\mbox{\epsfxsize10.0cm\epsffile[70 250 480 550]{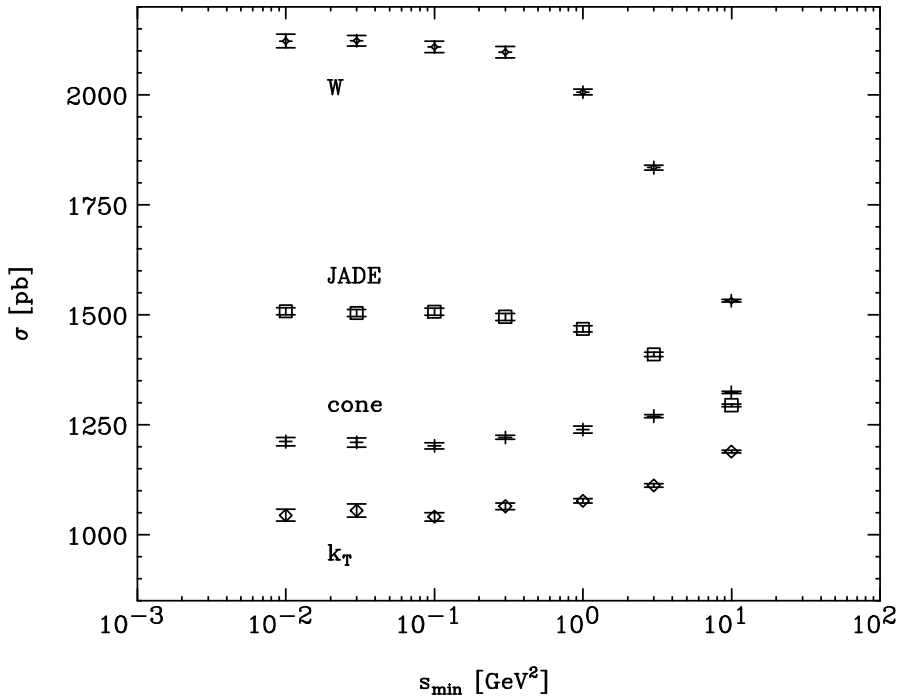}}
\end{picture}
\vspace*{4cm}
\caption{
Dependence of the inclusive two-jet cross section
in the $k_T$, cone, JADE, and the
$W$-scheme on $s_{min}$, the two-parton resolution parameter. 
Partons are recombined in the $E$-scheme. 
Error bars represent statistical errors of the Monte Carlo program. 
For the fairly soft jet definition criteria described in the text, 
$s_{min}$ independence is achieved for $s_{min}\lsim 0.1$~GeV$^2$. 
}\label{fig1}
\end{figure}

A powerful test of the numerical program is the $s_{min}$ independence of 
the NLO two jet cross sections for all jet algorithms.
Fig.~1 shows the inclusive dijet cross section as a 
function of $s_{min}$ for the four jet algorithms.
As mentioned before, $s_{min}$ is an arbitrary theoretical parameter 
and any measurable quantity should not depend on it. One observes
that for values smaller than 0.1~GeV$^2$ the results are indeed  
independent of $s_{min}$. The strong $s_{min}$ dependence of the NLO 
cross sections for larger values shows that the soft 
and collinear approximations used in the phase space region $s_{ij}<s_{min}$
are no longer valid, {\it i.e.} terms of ${\cal{O}}(s_{min})$ and
${\cal{O}}(s_{min}\ln s_{min})$ become important.
In general, one wants to choose $s_{min}$ as large as possible to avoid large
cancellations between the virtual+collinear+soft part ($s_{ij}<s_{min}$)
and the hard part of the phase space  ($s_{ij}>s_{min}$).
Note that factor 10 cancellations occur between the effective 2-parton and 
3-parton final states at the lowest $s_{min}$ values in Fig.~1 
and hence very high 
Monte Carlo statistics is required for these points. $s_{min}$ independence 
is achieved at and below $s_{min}=0.1$~GeV$^2$ and we choose this value for 
our further studies.

%All integrations over the hard phase space are done by Monte-Carlo
%techniques, which provides great flexibility to impose arbitrary
%cuts and to investigate different jet definition schemes. 

\section{Dijet Cross Sections in NLO}
\subsection{K-Factors and Recombination Scheme Dependence}
Table~\ref{table1} shows the importance of higher order corrections
and recombination scheme dependencies  \cite{bethke}
of the two jet cross sections
for the four jet algorithms.
While the higher order corrections and recombination scheme
dependencies in the cone
and $k_T$ schemes are small, very large corrections appear in the $W$-scheme. 
In addition, the large effective $K$-factor 
(defined as $K=\sigma_{NLO}/\sigma_{LO}$) 
of 2.04 (2.02) for the two-jet inclusive
(exclusive) cross section in the $W$-scheme depends strongly 
on the recombination scheme  which is used in the
clustering algorithm. 
Such large dependencies are subject to potentially large higher order
uncertainties, since the recombination dependence is only simulated at tree
level in the NLO calculation.

The large corrections and recombination scheme dependencies 
in particular in the $W$ scheme 
can partly be traced to large single jet masses 
(compared to their energy
in the parton center of mass frame).
As has been shown in \cite{mz1},
50 \% of the events in the NLO cross section for the $W$ scheme
(with the $E$ recombination scheme)
have a massive jet with $m/E > 0.44$, while substantially smaller 
values are found in the other jet schemes.
The very large median value of $m/E$ in the $W$-scheme implies that at NLO 
we are dealing with very different types of jets than at LO, and this 
difference accounts for the large $K$-factor.

In the JADE-algorithm the $K$-factor is reduced from 
1.48 in the $E$-scheme to 1.36 and 1.24 in the $E0$ and 
$P$-schemes\footnote{The NLO two jet cross sections for the $W$ or the JADE
scheme in Table I disagree with previous calculations \cite{wscheme2}.
The DISJET program\cite{disjet}, for example, gives a $K$-factor very 
close to unity for a phase space region which is very similar.}.
For the  cone ($k_T$) scheme this recombination scheme
dependence is reduced to the 3\% (10\%) level.

\begin{table}[h]
\caption{Two-jet cross sections in DIS at HERA. Results are given at LO and
NLO for the four jet definition schemes and acceptance cuts described in 
the text. The 2-jet inclusive cross section at NLO is given for three 
different recombination schemes.
}\label{table1}
\vspace{2mm}
\begin{tabular}{lccccc}
        \hspace{2.3cm}
     &  \mbox{2-jet }
     &  \mbox{2-jet exclusive}
     &  \mbox{2-jet inclusive}
     &  \mbox{2-jet inclusive} 
     &  \mbox{2-jet inclusive} \\
     &  \mbox{LO}
     &  \mbox{NLO} ($E$)
     &  \mbox{NLO} ($E$)
     &  \mbox{NLO} ($E0$)
     &  \mbox{NLO} ($P$)\\
\hline\\[-3mm]
\mbox{cone} & 1107~pb & 1047~pb & 1203~pb & 1232 pb  & 1208 pb \\
$k_T$       & 1067 pb & 946 pb  & 1038 pb & 1014 pb  & 944 pb \\
$W$         & 1020 pb & 2061 pb & 2082 pb & 1438 pb  & 1315 pb \\
\mbox{JADE} & 1020 pb & 1473 pb & 1507 pb & 1387 pb  & 1265 pb \\
\end{tabular}
\end{table}

\setlength{\unitlength}{0.7mm}
\begin{figure}[hbt]               \vspace*{-2cm}
\begin{picture}(150,165)(-30,1)
\mbox{\epsfxsize10.0cm\epsffile[78 222 480 650]{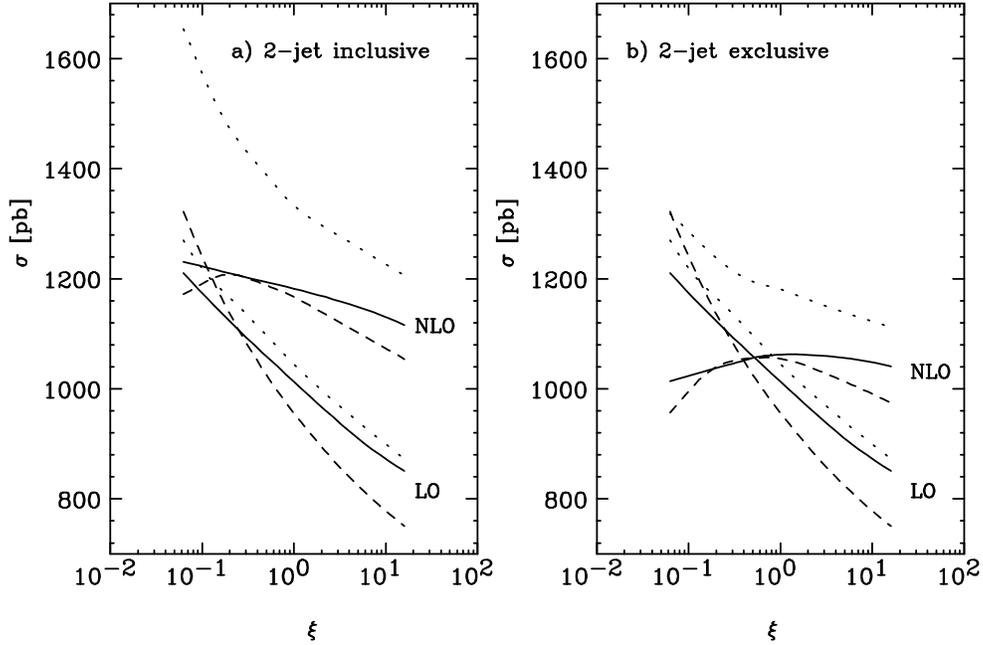}}
\end{picture}
\caption{
Dependence of a) the two-jet inclusive and b) the two-jet exclusive cross 
section in the cone scheme 
on the renormalization and factorization scale factor $\xi$.
The solid curves are for $\mu_R^2=\mu_F^2=\xi\;(\sum_i\;p_T^B(i))^2$, while for
the dashed curves only $\xi_R=\xi$ is varied but $\xi_F=1/4$ is fixed.
Choosing the photon virtuality as the basic scale yields the dotted curves,
which correspond to $\mu_R^2=\mu_F^2=\xi\;Q^2$. Results are shown 
for the LO (lower curves) and NLO calculations.
}\label{fig2}
\end{figure}

\subsection{Scale Dependence}

As mentioned before, the NLO corrections substantially reduce the 
the renormalization and 
factorization scale dependence which is present in the LO calculations 
and thus reliable cross section predictions in terms of $\alpha_s(m_Z)$
are made possible.
The scale dependence for  dijet cross sections in the cone scheme
is shown in Fig.~2. We have considered scales related to the 
scalar sum of the parton transverse momenta in the Breit frame,
$\sum_i \,p_T^B(i)$, and the virtuality $Q^2$ of the incident photon.
In Fig.~\ref{fig2} the dependence of the two-jet 
cross section, in the cone scheme, on the renormalization and 
factorization scale factors $\xi_R$ and $\xi_F$ is shown. For scales related 
to $\sum_i \,p_T^B(i)$ they are defined via
\begin{equation}
 \mu_R^2 = \xi_R\;(\sum_i \,p_T^B(i))^2\, ,
\hspace{1cm}
 \mu_F^2 = \xi_F\;(\sum_i \,p_T^B(i))^2\,.
\label{xidef}
\end{equation}
For the two-jet inclusive cross section of Fig.~\ref{fig2}a,
the LO variation by a factor 1.43 is reduced to a 10\%
variation at NLO when both scales are varied simultaneously 
over the plotted range (solid curves).
However, neither the LO nor the NLO curves show an extremum. 
The uncertainty from the variation of both scales  for the NLO two-jet
exclusive cross section in Fig.~2b (solid curves) is reduced to 5\%.
Furthermore, the two-jet exclusive cross section now has a 
maximum and is equal to the LO cross section for $\xi=0.5$.
Also shown is the $\xi=\xi_R$ dependence of LO and NLO cross
sections at fixed $\xi_F=1/4$ (dashed curves). In this case 
a maximum appears in the NLO inclusive and exclusive cross sections. 
However, the scale variation is stronger than in the $\xi=\xi_R=\xi_F$ case.

An alternative scale choice might be $\mu_R^2=\mu_F^2=\xi\,Q^2$.
The resulting $\xi$ dependence is shown as the dotted lines for
both the LO and NLO calculations.
At LO the two scale choices give qualitatively similar results.
However, with $\mu_R^2=\mu_F^2=\xi\,Q^2$, the scale dependence 
does not markedly  improve at NLO.
% within the kinematical range considered here. 
%In addition a sizable $K$-factor is found, with $K>1$ for small values
%of $Q^2$ and $K<1$ for very hard incident photons. 
We therefore use 
the jet transverse momenta in the Breit frame to set the scale
and fix $\xi_R=\xi_F=1/4$ in Eq.~(\ref{xidef}) for the 
following numerical results.
A careful study of the scale dependence and the choice
of the scale in the dijet cross section
is needed in order to extract a reliable value for 
$\alpha_s(M_z)$.

\section{Gluon Density Determination}
HERA opens  a  new window to measure the proton structure functions,
in particular the gluon distribution, in a completely new kinematic
region. The accessible range in the Bjorken-scaling variable $x$ can be 
extended considerably towards low $x$ 
compared to previous fixed target experiments.
Dijet production in DIS at HERA in principle allows for a more direct
measurement of the gluon density in the proton (via  $\gamma g \rightarrow
q\bar{q}$) than  an analysis of the structure function $F_2$.

For these studies we use the cone scheme as defined in section II.
The $Q^2$ range is lowered to $5<Q^2< 2500$ GeV$^2$ 
and the cut on the jet transverse momenta in the 
Breit frame  is increased to 5 GeV.
The LO (NLO) results are based on 
the LO (NLO) parton distributions from GRV \cite{grv} together with
the one-loop (two-loop) formula 
for the strong coupling constant.
With these parameters, one obtains 
2890 pb (2846 pb) for the LO (NLO) two jet exclusive cross section.

In order to investigate 
the feasibility of the parton density determination
at low $x$, Fig.~3a  shows the Bjorken $x$ distribution of the
two jet exclusive cross section in the cone scheme.
The gluon initiated subprocess clearly dominates the Compton process
for small $x$ in the LO predictions. The effective $K$-factor close to unity
for the total exclusive dijet cross section is a consequence of
compensating effects in the low $x$ ($K>$ 1) and high $x$ ($K<$1) regime.

\begin{figure}
\vspace*{-5cm}
\begin{picture}(150,165)(-30,1)
\mbox{\epsfxsize10.0cm\epsffile[78 222 480 650]{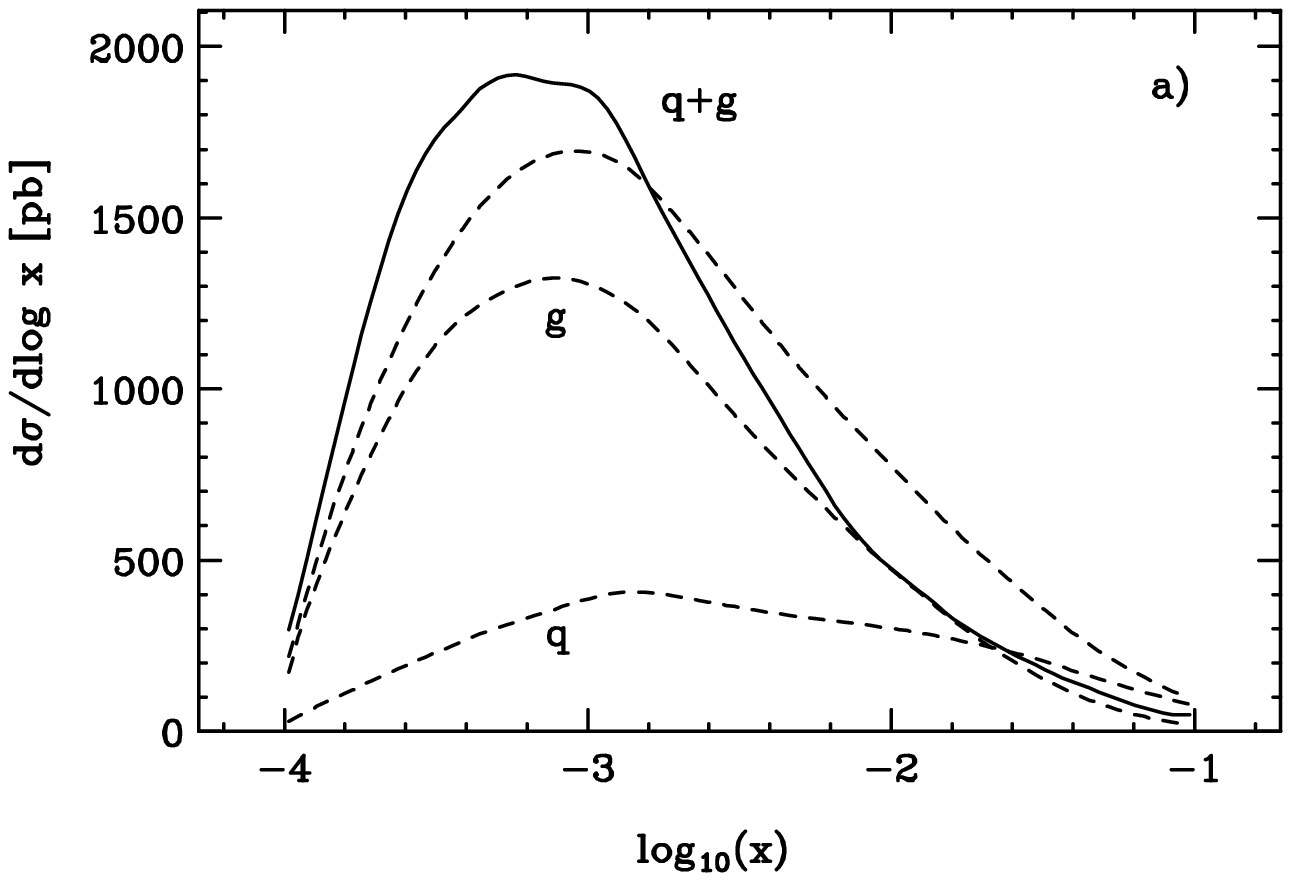}}
\end{picture}
\mbox{}\\[-5cm]
\begin{picture}(150,165)(-30,1)
\mbox{\epsfxsize10.0cm\epsffile[78 222 480 650]{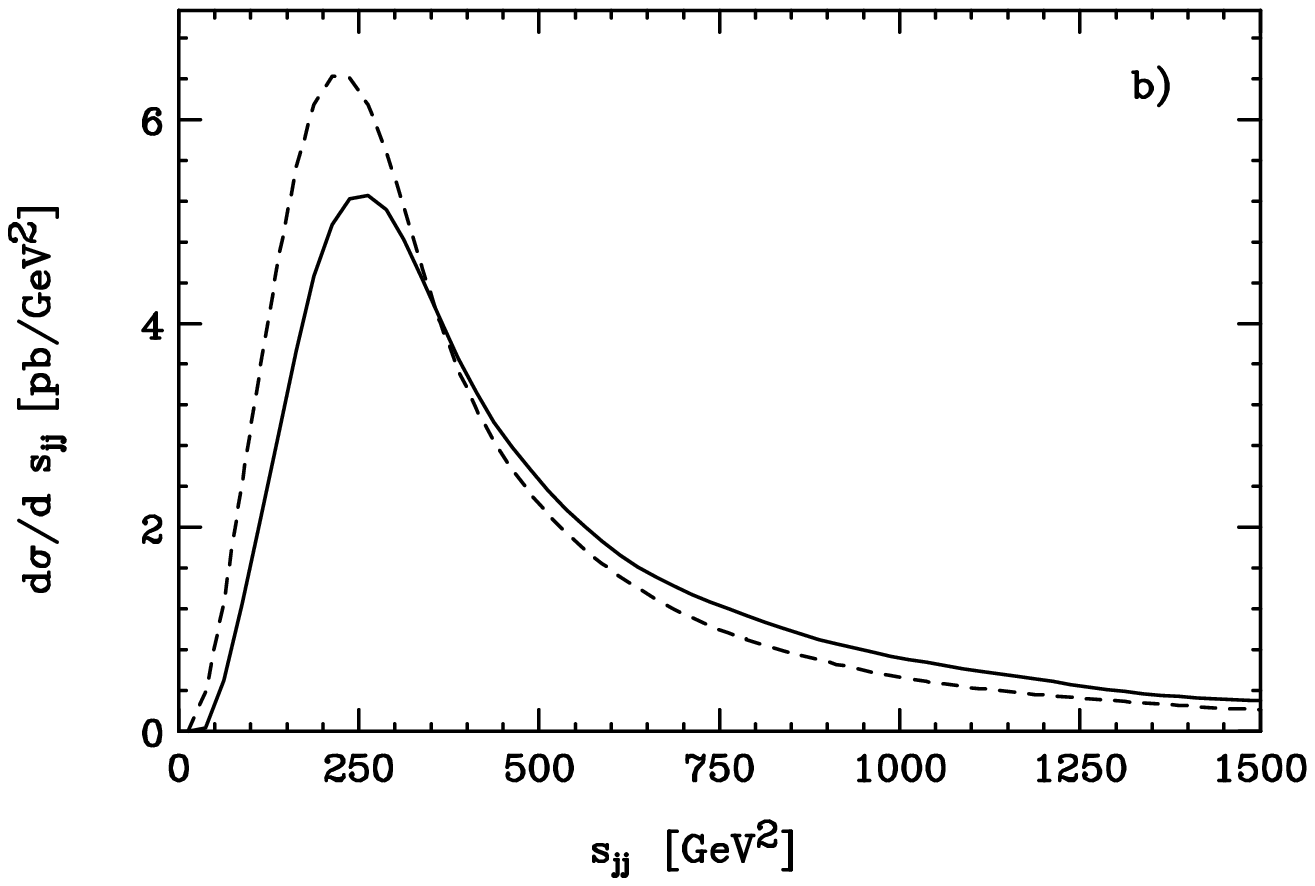}}
\end{picture}
\mbox{}\\[-5cm]
\begin{picture}(150,165)(-30,1)
\mbox{\epsfxsize10.0cm\epsffile[78 222 480 650]{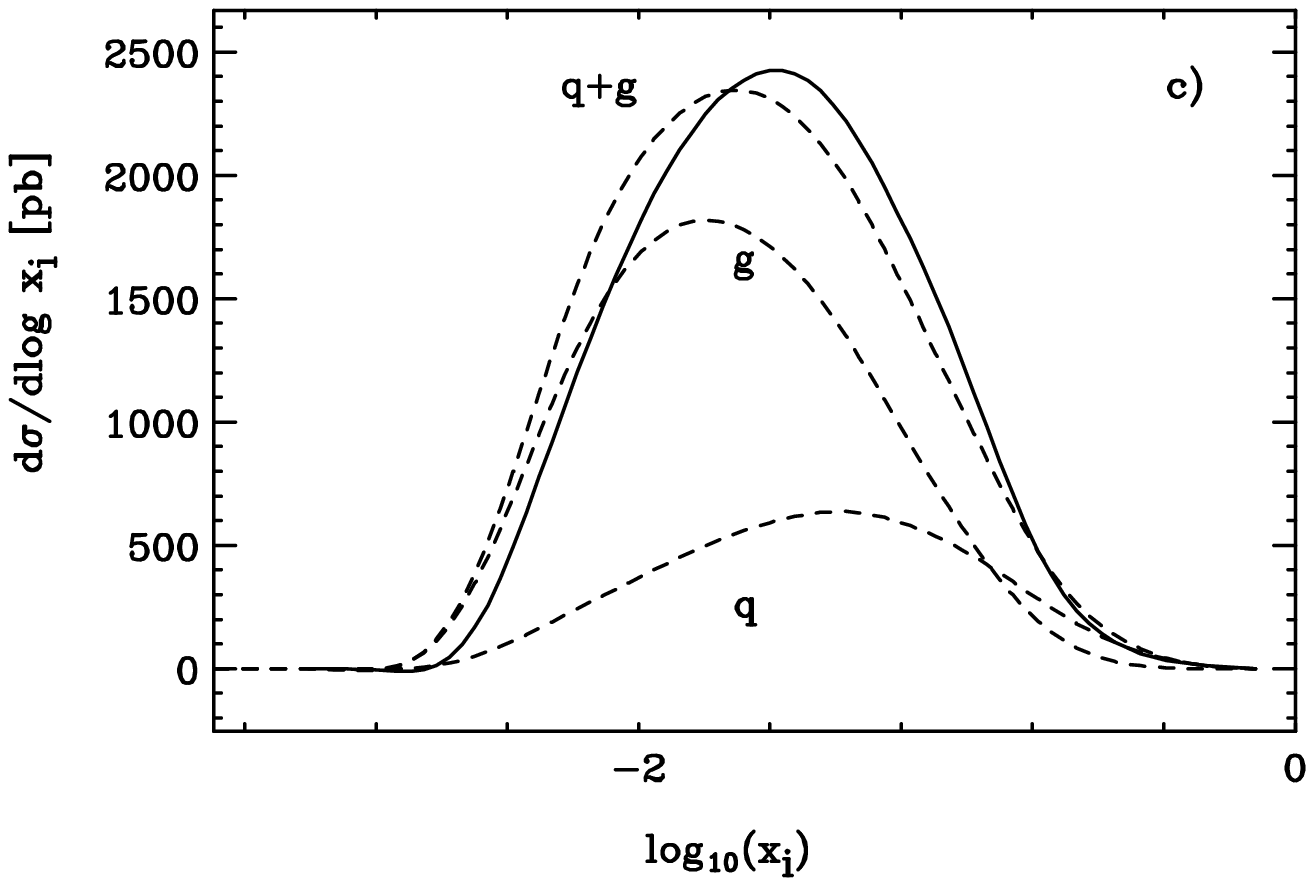}}
\end{picture}
\caption{
a) Dependence of the exclusive two-jet cross section
in the cone scheme on Bjorken $x$ for the
quark and gluon initiated subprocesses and for the 
sum. Both LO (dashed) and NLO (solid) results are shown;
b) Dijet invariant mass distribution in LO (dashed) and in NLO (solid);
c) Same as a) for the $x_i$ distribution, $x_i$ representing the 
momentum fraction of the incident parton at LO.
}
\end{figure}

For the isolation of parton structure functions we are interested in the 
fractional momentum $x_i$ of incoming parton $i$ ($i=q,g$), however, and in
dijet production $x$ and  $x_i$ differ substantially. Denoting as ${s_{jj}}$
the invariant mass squared of the produced dijet system, and considering 
two-jet exclusive events only, the two are related by
\begin{equation}
x_i = x \,\left(1+\frac{{s_{jj}}}{Q^2}\right)
\end{equation}
The $s_{jj}$ distribution of Fig.~3b exhibits rather large NLO
corrections as well. The invariant mass squared of the two jets is considerable
larger at NLO than at LO (the mean value of $s_{jj}$ rising to 570~GeV$^2$ 
at NLO from 470~GeV$^2$ at LO).

The NLO corrections to the $x$ and $s_{jj}$ distributions
have a compensating effect on the $x_i$ distribution in Fig.~3c, which shows
very similar shapes at LO and NLO. At LO a direct determination  of the gluon
density is possible from this distribution, after substraction of the 
calculated Compton subprocess. This simple picture is modified in NLO,
however, and the effects of Altarelli-Parisi splitting and low $p_T$ partons
need to be taken into account more carefully to determine the structure
functions at a well defined factorization scale $\mu_F$. We will further
investigate this problem
in a subsequent publication. One method to determine the gluon density
in NLO is presented in \cite{gluon}.

\section{Forward Jet Production in the Low $x$ Regime}
Deep inelastic scattering with a measured forward jet 
with relatively large momentum fraction $x_{jet}$
(in the proton direction) and $p_T^{2\,lab}(j)\approx Q^2$
is expected to provide 
sensitive information about the BFKL dynamics at low $x$
\cite{mueller,allen1}.
In this region there is not much phase space 
for DGLAP evolution with transverse momentum ordering,
whereas large effects are expected for BFKL evolution in $x$.
In particular, BFKL evolution is expected to substantially enhance cross
sections in the region $x<<x_{jet}$ \cite{mueller,allen1}.
In order to extract information on the $\ln(1/x)$
BFKL evolution, one needs to show that cross section results based on fixed 
order QCD with DGLAP evolution are not sufficient to describe the data. 
Clearly, next-to-leading order QCD corrections to the DGLAP predictions
are needed to make this comparison between experiment and theory.

In Table~\ref{table2} we show numerical results for the multi jet cross 
sections with (or without) a forward jet. 
The LO (NLO) results are based on 
the LO (NLO) parton distributions from GRV \cite{grv} together with
the one-loop (two-loop) formula 
for the strong coupling constant.
%The calculations use the parton 
%distribution functions  MRS D-' \cite{mrs}.
Kinematical cuts are imposed to closely model the H1 event 
selection\cite{deroeck}. More specifically, we require 
$Q^2>8~$GeV$^2$ , $x<0.004$,
$0.1 < y < 1$, an energy cut of $E(e^\prime)>11$~GeV on the scattered 
electron, and a cut on the pseudo-rapidity $\eta=-\ln\tan(\theta/2)$
of the scattered lepton of 
$ -2.868 < \eta(e^\prime)< -1.735$ 
(corresponding to $160^o < \theta(l^\prime) < 173.5^o$).
Jets are defined in the cone scheme (in the laboratory frame) with
$\Delta R = 1$ and $|\eta(j)|<3.5$.
We require a forward jet with $x_{jet}=p_z(j)/E_{P} > 0.05$,
$E(j)>25$ GeV, $0.5<p_T^2(j)/Q^2<4$,
and a cut on the pseudo-rapidity of 
$ 1.735< \eta(j)< 2.9$ 
(corresponding to $6.3^o < \theta(j) < 20^o$).
In addition all jets must have 
transverse momenta of at least  4 GeV in the lab frame
and 2 GeV in the Breit frame.

\begin{table}[t]
\caption{Forward jet cross sections in [pb] in DIS at HERA. 
}\label{table2}
\vspace{2mm}
\begin{tabular}{lccc}
        \hspace{0.8cm}
     &  \mbox{with  }
     &  \mbox{without  }
     &  \mbox{relative} \\
     &  \mbox{forward jet}
     &  \mbox{forward jet}
     &  \mbox{phase space}\\
\hline\\[-3mm]
\mbox{1 jet (LO)}   & 0    pb &  9026 pb   & 0\%     \\
\mbox{2 jet (LO)}   & 19.3 pb &  2219 pb   & 0.87\%  \\
\mbox{2 jet (NLO)}  & 68   pb &  2604 pb   & 2.61\%  \\
\mbox{3 jet (LO)}   & 30.1   pb & 450  pb   & 6.7\%   \\
\end{tabular}
\end{table}

The cross sections of Table~\ref{table2} demonstrate first of all that
the requirement of a forward jet with large longitudinal momentum fraction
($x_{jet}>0.05$) and restricted transverse momentum ($0.5<p_T^2(j)/Q^2<4$)
severely restricts the available phase space, in particular for low jet
multiplicities. The 1-jet exclusive cross section vanishes at LO, due to the
contradicting $x<0.004$ and $x_{jet}>0.05$ requirements. For $x<<x_{jet}$,
a high invariant mass hadronic system must be produced by the photon-parton
collision and this condition translates into 
\begin{equation}
2E(j)m_T\;e^{-y} \approx \hat{s}_{\gamma,parton} \approx 
Q^2\left({x_{jet}\over x}-1\right) >> Q^2\; ,
\end{equation}
where $m_T$ and $y$ are the transverse mass and rapidity of the 
partonic recoil system, respectively. Thus a recoil system with substantial
transverse momentum and/or invariant mass must be produced and this 
condition favors recoil systems composed out of at least two additional 
energetic partons. 

As a result 
one finds very large fixed order perturbative QCD corrections (compare
2 jet LO and NLO results with a forward jet in Table~\ref{table2}). 
In addition, the LO $({\cal{O}}(\alpha_s^2))$ 3-jet cross section 
is larger than the LO $({\cal{O}}(\alpha_s))$
2-jet cross section.
Thus, the forward jet  cross sections in Table~\ref{table2} are dominated
by the $({\cal{O}}(\alpha_s^2))$ matrix elements.
The effects
of BFKL evolution must be seen and isolated on top of these fixed order QCD
effects. We will analyze these effects in a subsequent publication.  

\section{Conclusions}

The calculation of NLO perturbative QCD corrections has received an enormous
boost with the advent of full NLO Monte Carlo programs 
\cite{giele1,giele2,jim}. For dijet production at HERA the NLO Monte Carlo
program MEPJET \cite{mz1} allows to study jet cross sections for arbitrary
jetalgorithms. Internal jet structure, parton/hadron recombination effects, 
and the effects of arbitrary acceptance cuts can now be simulated at the full
${\cal O}(\alpha_s^2)$ level.
We found large NLO effects for some jet definition schemes (in particular the
$W$-scheme) and cone and $k_T$ schemes appear better suited for precision
QCD tests. 

The extraction of gluon distribution functions is now supported by
a fully versatile NLO program. Preliminary studies show that large NLO 
corrections are present in the Bjorken $x$ distribution for dijet events,
while these effects are mitigated in the reconstructed Feynman $x$ ($x_i$)
distribution, thus aiding the reliable extraction of $g(x_i,\mu_F^2)$.

For the study of BFKL evolution by considering events with a forward 
``Mueller''-jet very large  
QCD corrections are found at ${\cal O}(\alpha_s^2)$.
These fixed order effects form an important background to the observation
of BFKL evolution at HERA. They can now be studied systematically and for
arbitrary jetalgorithms.

\acknowledgements
This research was supported by the University of Wisconsin 
Research Committee with funds granted by the Wisconsin Alumni Research 
Foundation and by the U.~S.~Department of Energy under Grant 
No.~DE-FG02-95ER40896. The work of E.~M. was supported in part  
by DFG Contract Ku 502/5-1.

%%%%%%%%%%%%%%%%%%%%% REFERENCES %%%%%%%%%%%%%%%%%%%%%%%%%%%%%%%%%%%%%%%%%%%%%%
%

%\end{thebibliography}
%

\end{document}